\documentclass[journal]{vgtc}         % review (journal style)
\ifpdf%                                % if we use pdflatex
  \pdfoutput=1\relax                   % create PDFs from pdfLaTeX
  \usepackage{graphicx}                % allow us to embed graphics files
  \DeclareGraphicsExtensions{.pdf,.png,.jpg,.jpeg} % for pdflatex we expect .pdf, .png, or .jpg files
\else%                                 % else we use pure latex
  \usepackage{graphicx}                % allow us to embed graphics files
  \DeclareGraphicsExtensions{.eps}     % for pure latex we expect eps files
\fi%

\graphicspath{{figures/}{pictures/}{images/}{./}} % where to search for the images

\usepackage[utf8]{inputenc}

\usepackage{microtype}                 % use micro-typography (slightly more compact, better to read)
\PassOptionsToPackage{warn}{textcomp}  % to address font issues with \textrightarrow
\usepackage{textcomp}                  % use better special symbols
\usepackage{mathptmx}                  % use matching math font
\usepackage{times}                     % we use Times as the main font
\usepackage{cite}                      % needed to automatically sort the references
\usepackage{tabu}                      % only used for the table example
\usepackage{booktabs}                  % only used for the table example
\usepackage{enumitem}
\setlist[description]{leftmargin=\parindent,labelindent=0pt}
\usepackage{scalerel}
\usepackage{ccicons}
\onlineid{1044}

\vgtccategory{Research}
\vgtcpapertype{algorithm/technique}

\title{Responsive Matrix Cells: A Focus+Context Approach for\\ Exploring and Editing Multivariate Graphs}

\author{Tom Horak$^{\ast}$, Philip Berger$^{\ast}$, Heidrun Schumann, Raimund Dachselt, Christian Tominski}

\authorfooter{
\item[$^{\ast}$] The first two authors contributed equally to this work.
\item
 T. Horak and R. Dachselt are with the Interactive Media Lab at Technische Universit\"{a}t Dresden. E-mail: \{horakt, dachselt\}@acm.org.
\item
 P. Berger, H. Schumann, C. Tominski are with the Inst. for Visual \& Analytic Computing at University of Rostock. E-mail: first.last@uni-rostock.de.
}

\shortauthortitle{Horak \MakeLowercase{\textit{et al.}}: Responsive Matrix Cells: A Focus+Context Approach for Exploring and Editing Multivariate Graphs}

\abstract{
Matrix visualizations are a useful tool to provide a general overview of a graph's structure.
For multivariate graphs, a remaining challenge is to cope with the attributes that are associated with nodes and edges.
Addressing this challenge, we propose \emph{responsive matrix cells} as a focus+context approach for embedding additional interactive views into a matrix.
Responsive matrix cells are local zoomable regions of interest that provide auxiliary data exploration and editing facilities for multivariate graphs.
They behave responsively by adapting their visual contents to the cell location, the available display space, and the user task.
Responsive matrix cells enable users to reveal details about the graph, compare node and edge attributes, and edit data values directly in a matrix without resorting to external views or tools.
We report the general design considerations for responsive matrix cells covering the visual and interactive means necessary to support a seamless data exploration and editing.
Responsive matrix cells have been implemented in a web-based prototype based on which we demonstrate the utility of our approach.
We describe a walk-through for the use case of analyzing a graph of soccer players and report on insights from a preliminary user feedback session.
}


\teaser{
  \centering
  \includegraphics[width=\linewidth]{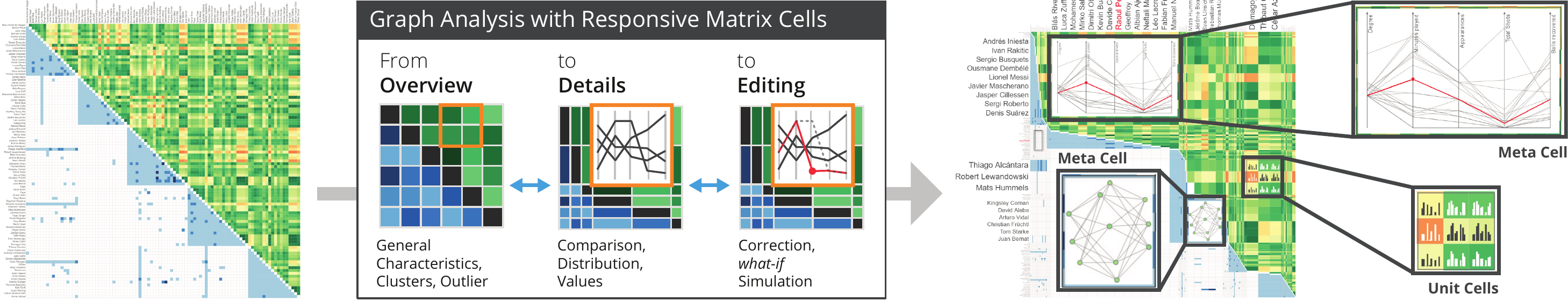}
  \caption{\emph{Responsive matrix cells} are a focus+context approach that provides details for a multivariate graph via embedded visualizations in a matrix representation and allows analysts to go from the overview in the matrix to details as well as editing within the cells. \ccby}
	\label{fig:teaser}
}

\vgtcinsertpkg

\definecolor{noteColor}{RGB}{173,110,27}
\definecolor{outdatedColor}{RGB}{143,143,143}
\definecolor{revisedColor}{RGB}{60,112,160}

\newcommand*{\inlinefig}[1]{\scalerel*{\includegraphics{#1}}{5}}
\begin{document}


\firstsection{Introduction}

\maketitle

%% \section{Introduction} %for journal use above \firstsection{..} instead

Multivariate graphs consist of nodes, edges, and multivariate data attributes.
An example would be a power grid, where power plants (the nodes) are characterized by quantitative attributes such as maximum capacity or current load.
Power lines (the edges) between plants can be characterized by attributes such as throughput or length.
In general, visualizing such graphs is challenging because multivariate attributes have to be visualized alongside the graph structure.

Typical tasks on multivariate graphs include gaining an overview of the graph structure (what is connected to what?), assessing the overall similarity of nodes (which power plants are alike?), studying the distribution of attribute values (what are the characteristics of plants in a sub-grid?), comparing nodes in detail (which plant produces less carbon dioxide?), and finding relations between attributes and the graph structure (are similar plants interconnected?)~\cite{Pretorius2014TasksMultivariateNetwork}.
In addition to these analysis-oriented objectives, it is becoming increasingly important to be able to edit or wrangle data~\cite{Baudel2006InformationVisualizationDirect, Kandel11Wrangling}.
Data editing can be necessary to correct erroneous data values (implausible power line throughput), and also to carry out \emph{what-if} analyses~\cite{Spence2001InformationVisualization} to test how data characteristics change when certain values are present in the data (would there be sufficient energy when reducing the capacity of some power plants?).

Solving the outlined analysis tasks typically requires an interplay of several visual representations~\cite{Kerren14MultivariateGraphVis, Nobre2019StateArtVisualizing}. On top of that, data editing usually requires external tools~\cite{Batch18Gap}.
However, switching between different visualization views and external tools may disrupt the data exploration and editing workflow.
Therefore, our goal is to enable users to go \emph{seamlessly from overview to detail to editing}.

To this end, we propose \emph{responsive matrix cells} (RMCs) as a matrix-based focus+context approach for integrated graph exploration and editing.
As shown in \autoref{fig:teaser}, a custom matrix visualization provides the overview of graph structure and multivariate data attributes.
RMCs act as local zoom areas that can be expanded dynamically in a focus+context fashion to embed additional interactive views.
These views then facilitate the analysis of details and the editing of data values directly within the matrix.
For example, substructures can be analyzed in detail with miniature node-link diagrams, attributes can be represented as small bar charts, or two nodes be compared directly via star plots.
Editing tasks can be performed directly in these auxiliary views, 
as soon as they have been zoomed sufficiently to allow for direct manipulation.
All this is possible within the overview matrix and without disruptive switches to external tools.
Our RMCs behave responsively by adapting their contents to where they are located in the matrix and how much display space is at their disposal. 
Moreover, users can adapt RMCs as needed for the task at hand.

In sum, our contributions are the following:
(1) \emph{responsive matrix cells} as a novel focus+context approach for exploring and editing multivariate graphs in an integrated fashion;
(2) \emph{design considerations} for embedding responsive visualizations into a matrix representation;
(3) \emph{interaction strategies} for a fluid and seamless analysis and editing of the data; and
(4) a \emph{web-based prototype} demonstrating the feasibility of our concepts, which is illustrated by means of a walk-through of a real-world dataset and by insights from a user feedback session.

\section{Related Work}

Related work is manifold.
We consider multivariate graph visualization, presentation techniques, as well as interaction and editing techniques for graphs.

\subsection{Multivariate Graph Visualizations}

Several approaches exist for visualizing multivariate graphs, with the majority of them being based on node-link diagrams or matrix visualizations~\cite{Kerren14MultivariateGraphVis, Nobre2019StateArtVisualizing}.
While node-link diagrams can be considered the default visualization for graphs, their layout can quickly get confusing, particularly when encoding additional data attributes. In contrast, matrix visualizations feature a clear and predictable layout suitable for providing an overview, even for dense graphs.

In order to make all aspects of a multivariate graph visually accessible, the aforementioned base visualizations must be extended.
This can be done by incorporating additional views~\cite{Kerzner2017GraffinityVisualizingConnectivity, Nobre2019JuniperTree+TableApproach}, embedding additional visual encodings~\cite{Major2019GraphicleExploringUnits, Elzen2014MultivariateNetworkExploration}, or laying out the graph based on its attributes~\cite{Wattenberg2006VisualExplorationMultivariate, Wu2008VisualizingMultivariateNetworks}.
While incorporating additional views makes it easier to encode more information, such solutions introduce a discontinuity between identifying regions of interest in one view and analyzing the actual details in another view.
As a result, relating information across views can impose a higher mental demand to the analyst.
Embedding additional visual encodings and varying the layout can avoid this, but it is usually only possible to visualize attributes in an abstract or aggregated form.
Therefore, existing solutions often favor one data aspect over another~\cite{Nobre2019StateArtVisualizing} or are geared towards specific analysis tasks~\cite{Pretorius2014TasksMultivariateNetwork}.

Our approach uses a matrix as the central graphical arrangement.
Matrices encode the presence of edges (or edge weights) in a tabular layout and are in general well-suited for visual graph analysis~\cite{Ghoniem2005ReadabilityGraphsUsing, Okoe2019NodeLinkAdjacency}.
Typical techniques for representing the edge attributes are color-coding and also small glyph-like visualizations placed directly in the matrix cells~\cite{Elmqvist2008ZAMEInteractiveLarge, Yi2010TimeMatrixAnalyzingTemporal}.
As matrices do not explicitly represent the graph nodes, additional means are required to visualize node attributes.
Prior research has assessed that a juxtaposed attribute table is a suitable solution~\cite{Berger2019VisuallyExploringRelations, Nobre2020EvaluatingMultivariateNetwork}.
An alternative is to calculate a pairwise attribute-based similarity measure for nodes and visualize it in one half of the matrix (divided by the diagonal), while the other half still encodes the edges~\cite{Berger2019VisuallyExploringRelations}.
This creates an overview of structural and attribute-based characteristics, enabling users to see, for example, whether nodes being similar with respect to their attributes are also connected by edges. We will use such a divided matrix design for our approach.

A disadvantage of matrices is their quadratic space complexity, which makes visualizing larger graphs demanding~\cite{Abello02GigaGraphs}.
Moreover, matrices are not very well suited for path-related tasks~\cite{Nobre2019StateArtVisualizing, Nobre2020EvaluatingMultivariateNetwork, Wong2013VisualMatrixClustering}.
A promising approach to mitigate these issues is to combine matrix and node-link representations, as in hierarchical graph maps~\cite{Abello02GigaGraphs} or NodeTrix~\cite{Henry2007NodeTrixHybridVisualization}.
Here, parts of the matrix are replaced with a node-link representation or vice versa for showing regions of interest in an alternative way.
Such local replacements and adaptations within the display are also part of general presentation techniques, as discussed next.

\subsection{Presentation Techniques}

Temporary local adaptations of a visual representation can help reveal details for regions of interest while the global context is preserved.
Focus+context techniques often apply a local zoom effect while maintaining the overall visualization dimensions.
Examples of focus+context techniques are bifocal displays~\cite{Apperley1982BifocalDisplayTechnique}, fisheye views~\cite{Furnas1986GeneralizedFisheyeViews, Rauschenbach2001Generalrectangularfisheye}, rubber-sheet navigation~\cite{Sarkar93Rubbersheet}, the table lens~\cite{Rao1994TableLensMerging}, the date lens~\cite{Bederson2004DateLensFisheyeCalendar}, or M{\'{e}}lange~\cite{Elmqvist2008MelangeSpaceFolding}.

Focus+context is not limited to geometrical scaling.
Semantic zooming can dynamically alter the layout or the very encoding of the focused parts of a visualization~\cite{Perlin1993PadAlternativeApproach}. Examples would be to change the type of chart embedded into the cells of a table lens~\cite{McLachlan08LiveRAC} or to show meta-nodes for clusters when zoomed out and to automatically expand the clusters to reveal their affiliated nodes when zooming in~\cite{Abello2006ASKGraphViewLarge, Shi2009HiMapAdaptivevisualization}.

Similar to focus+context techniques, magic lenses are lightweight tools that fluidly integrate a transient lens effect into the visualization~\cite{Tominski2016InteractiveLensesVisualization,Kister2016MultiLensFluentInteraction}.
In the context of graph visualization, lenses can, e.g., reduce clutter by filtering edges or generate local neighborhood overviews by adapting the layout~\cite{Tominski09CGV}. The latter one is also possible with our RMCs.
Similar to lenses, \emph{in situ visualization} allows users to interactively mark a region in a base visualization for which a different nested visualization is shown~\cite{Hadlak11InSitu}.

When considering the nesting of views to provide alternative representations locally on demand, the embedded visualizations have to face specific layout restrictions~\cite{Javed2012ExploringDesignSpace}.
For example, when embedding charts in table cells as in LiveRAC~\cite{McLachlan08LiveRAC} or glyphs in a matrix as in ZAME~\cite{Elmqvist2008ZAMEInteractiveLarge} or TimeCells~\cite{Yi2010TimeMatrixAnalyzingTemporal}, the available space is severely limited.
Depending on the application and the user's tasks, different visual encodings for such embedded or micro visualizations are possible~\cite{Beck2017WordSizedGraphics, Fuchs2017SystematicReviewExperimental, Tufte2006BeautifulEvidence}.

In the context of focus+context and semantic zooming, space constraints are more relaxed because users can freely define and change the zoom level and the dimensions of the focus region.
This makes it possible to add details to the visualization (e.g., labels, axes, or guides) or to switch to increasingly detailed visualization metaphors~\cite{MatkovicProcessvisualizationlevels, McLachlan08LiveRAC}.

Ideally, embedded visualizations are responsive, that is, they are able to adapt themselves automatically to external contextual requirements.
The notion of responsive visualizations has mostly been discussed in the context of mobile visualization~\cite{Lee2018DataVisualizationMobile, Choe2019MobileDataVisualization}.
Recent research suggests that the design of a responsive visualization should consider layout, data density, and interaction-related aspects~\cite{Andrews2017ResponsiveDataVisualisation, Andrews2018ResponsiveVisualisation, Hoffswell20Responsive}.
In the context of our RMC approach, similar strategies for responsiveness have to be taken into account.
Additionally, we consider whether the task of the user is to explore the data or to edit them.

\subsection{Interacting \& Editing in Graph Visualizations}

In general, interaction plays an important role for exploring multivariate graphs~\cite{Wybrow2014InteractionVisualizationMultivariate}.
The literature suggests that interaction can take place at different levels, including view-level interactions (e.g., brushing and linking), visual-structure interactions (e.g., selections), and data-level interactions (e.g., inserting or deleting edges).
Making selections in graphs or filtering nodes and edges are a fundamental operations~\cite{McGuffin09GraphInteraction, Tominski09CGV}.
A key interaction for matrix visualizations would be to re-order the node rows and columns to reveal different types of patterns~\cite{Perin14Matrices, Behrisch16MatrixOrdering}.

Interaction in graph visualizations is not limited to mere selections or adjustments of the visual representation.
Interaction is also relevant in the interplay of graph exploration and graph editing~\cite{Gladisch2015MappingTasksInteractions}.
Following Baudel's direct manipulation\footnote{Baudel's direct manipulation regards the direct editing of data values and is not to be mistaken for the classic notion of direct manipulation\cite{Shneiderman83DirectManipulation}.} principle~\cite{Baudel2006InformationVisualizationDirect}, previous work has proposed to edit node attributes by moving the nodes in a 2D-coordinate system with an overlaid node-link diagram~\cite{Eichner2016DirectVisualEditing}.
For editing a graph's structure, specialized lens tools can be employed~\cite{Gladisch2014SemiAutomaticEditing}.
Specifically for matrix visualizations, interactive editing approaches focus around adding or removing edges by (un)marking the corresponding matrix cells~\cite{Gladisch2015usingMatrixVisualizations,  Kister2017GraSpCombiningSpatially}.
More elaborate and integrated approaches, for example, for editing specific attributes of both nodes and edges, remain under-explored so far.

\subsection{Open Challenges}

Overall, it remains challenging to visually explore and also edit multivariate graphs.
To avoid unwanted attention switches and increased screen space demands, focus+context and semantic zooming have already been applied to data tables~\cite{Rao1994TableLensMerging,McLachlan08LiveRAC} and matrices~\cite{Abello02GigaGraphs, Elmqvist2008ZAMEInteractiveLarge, Yi2010TimeMatrixAnalyzingTemporal}.
However, the existing techniques are typically tailored to showing one specific data aspect of their respective data set.
Our approach is flexible and can show multivariate attributes as well as structural aspects of graphs on demand.
Moreover, our approach goes beyond focusing on individual matrix cells.
A responsive matrix cell can also span a sub-matrix and present the associated data in an aggregated form.
This is typically not possible with existing techniques, which prohibits insights into aggregated subsets of the data.
Finally, to our knowledge, none of the existing approaches considers data editing as an important task complementary to data exploration.

We aim to narrow this gap with an integrated approach that offers a promising novel way for multi-faceted exploration of multivariate graphs, together with the possibility to edit the data where necessary without losing the overall analysis context.

\section{Responsive Matrix Cells}

We propose \emph{responsive matrix cells} (RMCs) as a flexible focus+context approach to embed responsive visualizations into a matrix, more specifically, either into individual cells (unit cells) or across cohesive sub-matrices (meta cells).
In this section, we discuss the requirements for RMCs, provide an overview of the general approach, and elaborate on the visual design of RMCs.
The interaction facilities of RMCs will be described in detail later in \autoref{sec:overview-to-detail-to-editing}.

\subsection{Requirements}

Based on an analysis of the characteristics of multivariate graphs and the associated tasks~\cite{Kerren14MultivariateGraphVis}, we identified the following application-agnostic requirements for our approach.

\begin{description}
\itemsep0em 
    \item[R1: Provide overview.] Our approach must provide an overview of both graph structure and multivariate attributes, enabling analysts to spot general patterns (e.g., cliques or clusters), potential outliers, and possible relations between structure and attributes (e.g., similar nodes are connected).
    \item[R2: Allow access to details.] For selected regions of interest, it must be possible to access details to refine and complement the findings made with the overview. This includes identifying specific attribute values and comparing nodes or edges for concrete differences.
    \item[R3: Enable direct editing.] Editing should be possible directly in the visualization to allow users to quickly correct erroneous data or test \emph{what-if} scenarios while observing the resulting changes on the fly.
\end{description}

These requirements are concerned with \emph{what} we want to achieve.
On top of that, we address an additional requirement centered on \emph{how} we want to achieve R1--R3:

\begin{description}
\itemsep0em 
    \item[R4: Strive for a fully integrated approach.] All aspects inherent in multivariate graphs should be shown in an integrated visualization that supports data exploration and data editing. The integrated approach is to support smooth dynamic workflows and reduce inconvenient attention switches between different tools.
\end{description}

By following R4, we aim to utilize the known advantages of integrating focus within context~\cite{Cockburn08Survey}, the visual information seeking mantra~\cite{Shneiderman1996}, and direct editing~\cite{Baudel2006InformationVisualizationDirect}.

\subsection{Approach Overview}

\begin{figure}
    \centering
    \includegraphics[width=\columnwidth]{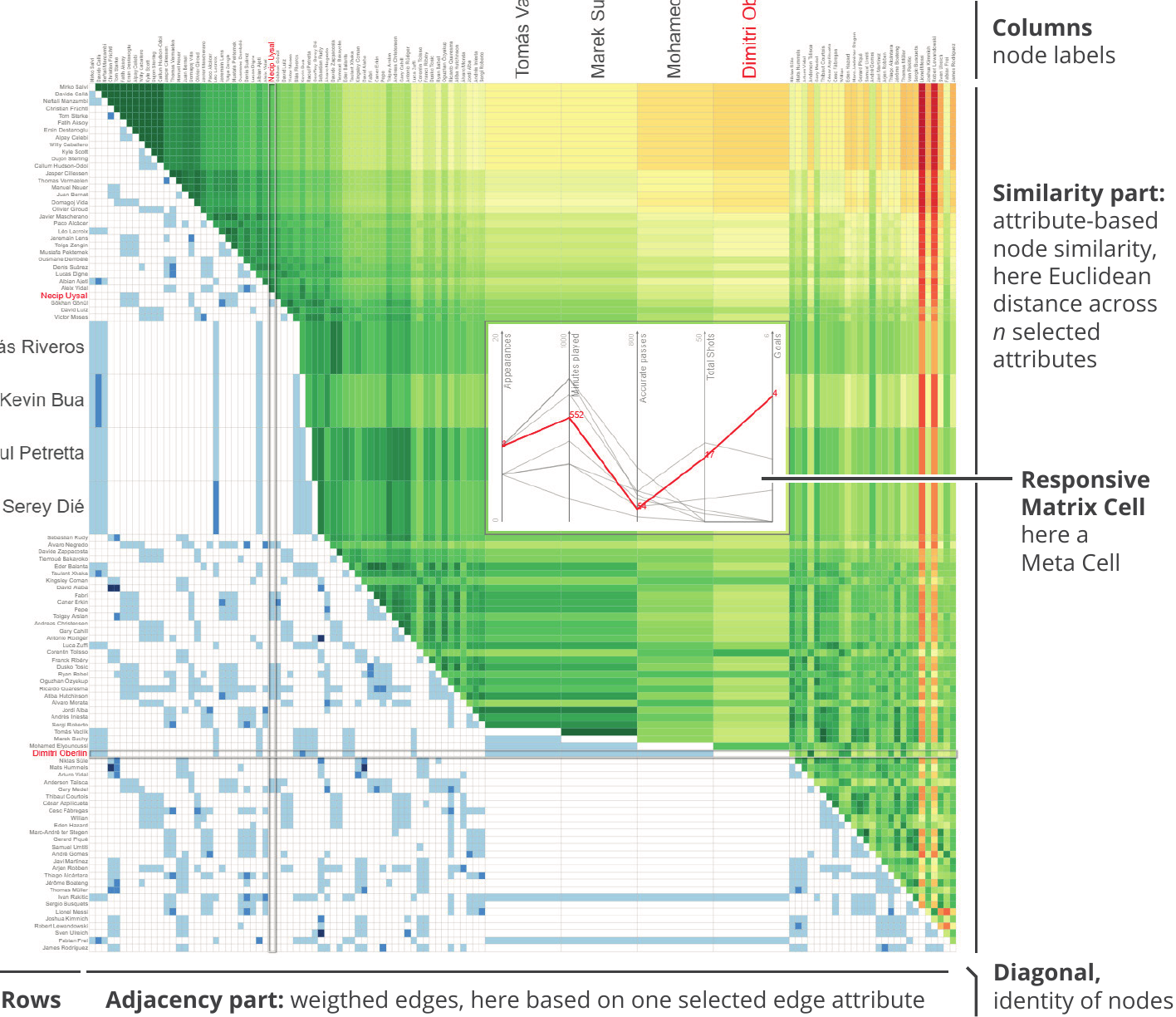}
    \vspace{-15pt}
    \caption{Matrix, where the lower triangular half visualizes the weighted edges of a graph, whereas the upper triangular half displays the pairwise similarity of nodes with respect to their multivariate attributes. A responsive matrix cell (RMC) is embedded into the matrix. \ccby}
    \label{fig:matrix-details}
\end{figure}

The core idea of the RMC approach is illustrated in \autoref{fig:matrix-details}: A special matrix visualization delivers the overview, while responsive matrix cells embedded into the matrix provide details in various ways and allow users to perform edit operations.

The basis for the overview is a customized matrix visualization~\cite{Berger2019VisuallyExploringRelations}.
As depicted in \autoref{fig:matrix-details}, it shows adjacency information and node similarity at the same time.
As for regular adjacency matrices, rows and columns correspond to the set of nodes.
The lower-left triangular half of the matrix visualizes the presence of edges and color-codes a selected edge attribute.
Yet, the upper-right triangular part of the matrix shows different information.
It color-codes pairwise node similarity as computed based on node attributes.
This custom matrix allows users to recognize structural clusters (e.g., hub nodes, cliques, bi-cliques), groups with similar attribute values, and outliers in general (R1).
However, as the color-coding visualizes only a single piece of information (i.e., attribute value or node similarity) per cell, multivariate details of edges and nodes are not visible.

To access details and additional functionality, users can initiate responsive matrix cells (RMCs) within the overview matrix.
More specifically, users create RMCs either for individual matrix cells (unit cells) or for sub-matrices (meta cells) and scale them up in a focus+context fashion as shown in \autoref{fig:matrix-details}.
The gained display space is used to embed interactive views that enable users to see and compare details of the data (R2).
Additionally, editing facilities are provided when RMCs are shown at a sufficient size (R3).
This minimizes interruptions of the analysis workflow as users no longer need to resort to external editing tools (R4).

RMCs reveal details and functionality to the analyst in a responsive way.
The responsive behavior of RMCs is a key feature of our approach.
At the core, an embedded RMC visualization adapts to:

\begin{enumerate}
    \itemsep0em 
    \item the origin where the RMC has been created,
    \item the space being available for the RMC, and
    \item the task (i.e., explore, compare, edit) of the analyst.
\end{enumerate}

There are different design choices for making RMCs responsive.
We will primarily be concerned with \emph{what} additional information can be shown \emph{where} in the matrix, and \emph{how} the information can be visualized specifically (\autoref{fig:rmc-overview}).
The what, where, and how will be detailed in the remainder of this section.

The general possibility to embed visualizations into a matrix is only one side of RMCs.
The other side is concerned with the interactive interface required to enable users to utilize overview, detail views and editing facilities smoothly, which will be the topic of \autoref{sec:overview-to-detail-to-editing}.

\subsection{What can be Shown?}

Multivariate graphs consist of two types of \emph{objects}, nodes and edges, where each object can have several attribute values.
By having a matrix with an adjacency part (lower-left) and a similarity part (upper-right), one half of the matrix is primarily focused on the edges, while the other half is focused on the nodes, more specifically on how two given nodes compare.
This distinction is crucial to understand \emph{what} information is shown in RMCs.

As indicated in \autoref{fig:rmc-overview}, an RMC being located in the adjacency part (blue) will show information about the edges associated with the underlying matrix cells, while an RMC in the similarity part (green) will show information about the nodes associated with the corresponding matrix rows and columns.

An RMC may span a single matrix cell, in which case it either represents a single edge (adjacency part), a single node (diagonal), or a pair of two nodes (similarity part).
Such RMCs allow analysts to study the details of individual nodes and edges or conduct a 1:1 comparison of two nodes.
An RMC may also cover an $i \times j$ sub-matrix with $m = i \cdot j$ cells, which means it represents either a group of $n \leq i \cdot j$ edges or a group of $n \leq i + j$ nodes.
For such groups of objects, analysts might be interested in studying individual objects as indicated before, but also in investigating the group characteristics as a whole, including the distribution of attribute values or structural aspects of the group's induced sub-graph.

In sum, RMCs support three types of information representation: representations for \emph{1 object} to show its details, for \emph{2 objects} to directly compare them, or for \emph{$n$ objects} to convey group properties.

\subsection{Where will Information be Shown?}

The question of where detail information will be shown depends on a user-specified region of interest (RoI).
If the user is interested in an individual edge or an individual pair of nodes, the RoI consist of only a single cell.
In that case, a single visualization is embedded into the cell of interest.
We call such cells \emph{unit cells}.

\begin{figure}[t]
    \centering
    \includegraphics[width=\columnwidth]{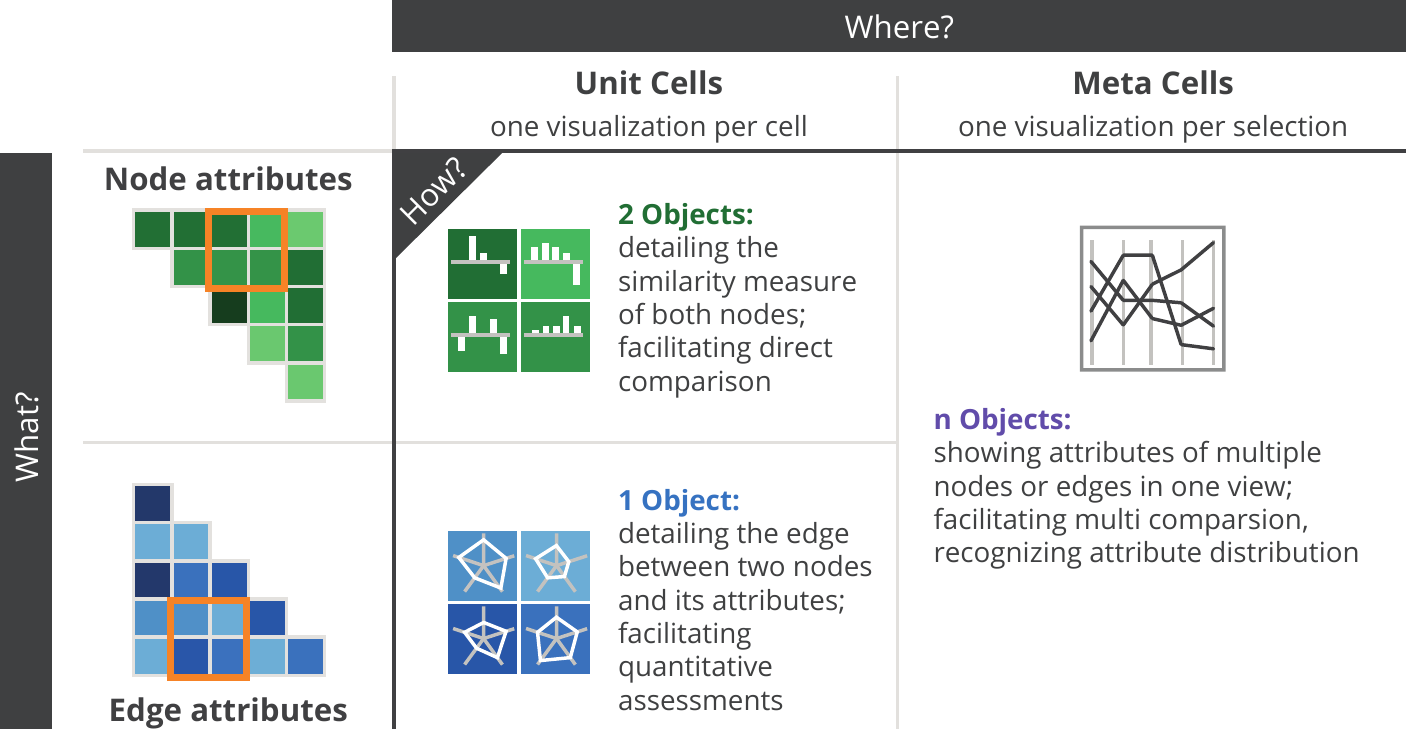}
    \caption{Responsive matrix cells are characterized by \emph{what} they show, \emph{where} they show it, and \emph{how} they show it. The \emph{what} and the \emph{where} define the context for the \emph{how}. \ccby}
    \vspace{-10pt}
    \label{fig:rmc-overview}
\end{figure}

When the RoI is defined as an  $i \times j$  sub-matrix, it could mean the user wants the details for (a) the individual objects covered or (b) the group comprised of the objects.
For case (a), multiple unit cells are created so that there is one embedded visualization for each cell of the sub-matrix.
In other words, the cells of the sub-matrix are treated individually as units, similar to \emph{small multiples}~\cite{Tufte83VisualDisplay}.
For case (b), the sub-matrix is treated as a whole and a single visualization is embedded into it.
We can also say that the RoI is subsumed into an aggregated \emph{meta cell} being concerned with the data as a group.
\autoref{fig:rmc-overview} illustrates that unit cells provide visualizations detailing $1$ or $2$ objects, whereas a meta cell provides the details for $n$ objects in a single visualization.

Unit cells and meta cells differ in their characteristics, which also has consequences for the embedded visualizations.
Unit cells generally start in the square aspect ratio of the underlying matrix cells.
When unit cells are generated for a sub-matrix, a visualization is placed in each cell.
As these visualizations have to share the available display space, they initially cover only a few pixels.
Therefore, unit cells typically require zooming before further details are revealed.
\autoref{fig:design_unit-cells} depicts possible unit cell designs.

Meta cells span multiple underlying matrix cells and therefore start at a larger size than unit cells.
Yet, as illustrated in \autoref{fig:design_meta-cells}, no assumptions can be made about a meta cell's aspect ratio as it depends on the shape of the RoI defined by the analyst.
Consequently, the visualizations embedded into meta cells must cope with varying aspect ratios.
Next, we discuss the design of embedded responsive visualizations in detail.

\subsection{How is Information Shown?}
\label{subsect:rmc-visualizations}

General design guidelines for embedding responsive visualizations do not yet exist in the literature.
Here, we primarily discuss what it takes to make visualizations responsive and illustrate them with selected examples.
Our discussion focuses on (i) how the visualizations scale and respond, and (ii) what information they can represent.

\subsubsection{Making Visualizations Responsive}

In our case, the embedded responsive visualizations must be able to communicate the characteristics of one or two objects for unit cells, and of $n$ objects for meta cells (\autoref{fig:rmc-overview}).
Depending on the number of objects, the visualizations should facilitate \emph{object visibility} or \emph{attribute visibility}~\cite{Spence2001InformationVisualization}.
The focus can be on representing data attributes or supporting comparison tasks (R2).
As indicated above, responsive visualizations must also be compatible with different aspect ratios.

Most importantly for our focus+context approach, the visualizations must be able to work at different sizes.
Ideally, details are conveyed already at sizes of a few pixels.
When additional space becomes available, it should be used efficiently by adding more and more details, not only geometrically, but also semantically~\cite{Perlin1993PadAlternativeApproach, MatkovicProcessvisualizationlevels, McLachlan08LiveRAC}.

For our RMCs, we consider four major levels of detail (LoD) that represent important breakpoints when increasing the cell size:

\begin{enumerate}
    \itemsep0em 
    \item \emph{Pixel} level with color-coding only,
    \item \emph{Miniature} level with a minimal version of the visualization,
    \item \emph{Compact} level showing first labels or values, and
    \item \emph{Medium} level showing more labels and details.
\end{enumerate}

Note that the medium level is not meant as a maximum, since cells can be increased even further and more details can be added.
Also, we refrain from defining exact pixel-based values for these sizes because the specific thresholds for showing additional details depend on the visualization (e.g., how space-efficient the visualization is), the used device (e.g., what resolution and pixel density is offered), and preferences of the user (e.g., details as soon as possible vs. abstraction as early as possible).

A general concern though is to help users maintain their mental map as the LoD changes.
To this end, we propose to preserve the original matrix cell's color-coding as the background color at the miniature size or as the border color for the larger sizes as illustrated in \autoref{fig:design_unit-cells}.
Maintaining the color as a visual residue can make it easier to keep track of specific cells and to recall why they seemed of interest in the first place (e.g., dark encoding, light encoding, similar encoding).
Yet, when used in the background, the color can potentially compromise the contrast in the embedded visualizations.
Therefore, miniature visualizations render their marks using a contrast color (e.g., white or dark gray) that depends on the luminance of the background.
This way, we can guarantee a sufficient separation of background and visualization.

Complementing the aforementioned general design aspects, we next discuss specific design considerations for visualizing the multivariate attributes of nodes and edges.
Representations of structural aspects and multi-faceted data aspects will be discussed later in \autoref{subsect:furtheraspects}.

\subsubsection{Designs for Multivariate Aspects}
\label{subsect:multivariateaspects}

\begin{figure}[t]
    \centering
    \includegraphics[width=\columnwidth]{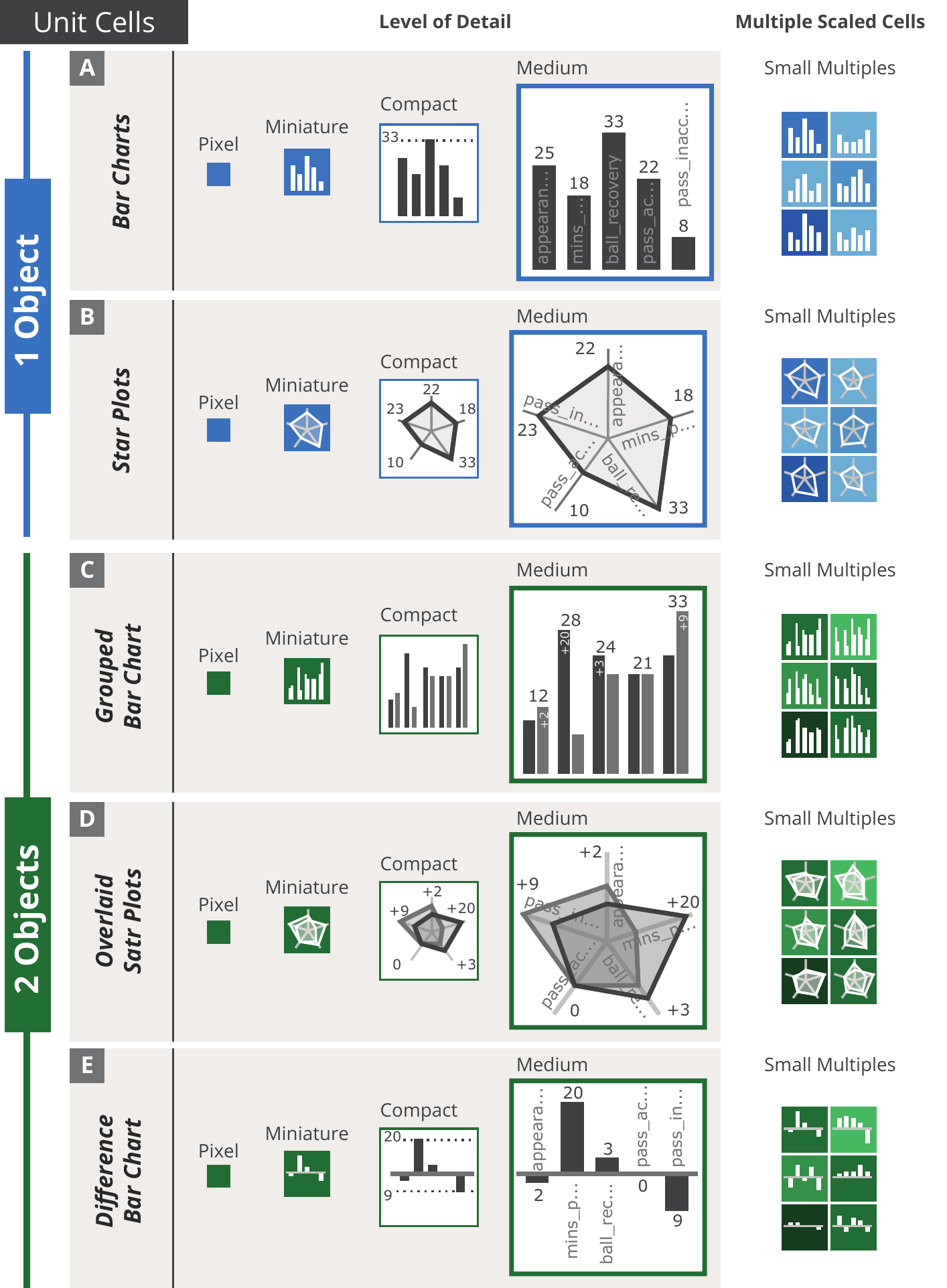}
    \vspace{-15pt}
	\caption{Visualizations in unit cells have to represent 1 or 2 objects. Variants of bar charts and star plots are suitable for being embedded into unit cells. Depending on the available display space, different levels of detail can be offered. Unit cells also work as small multiples. \ccby}
	\vspace{-10pt}
	\label{fig:design_unit-cells}
\end{figure}

\begin{figure*}[t]
    \centering
    \includegraphics[width=\textwidth]{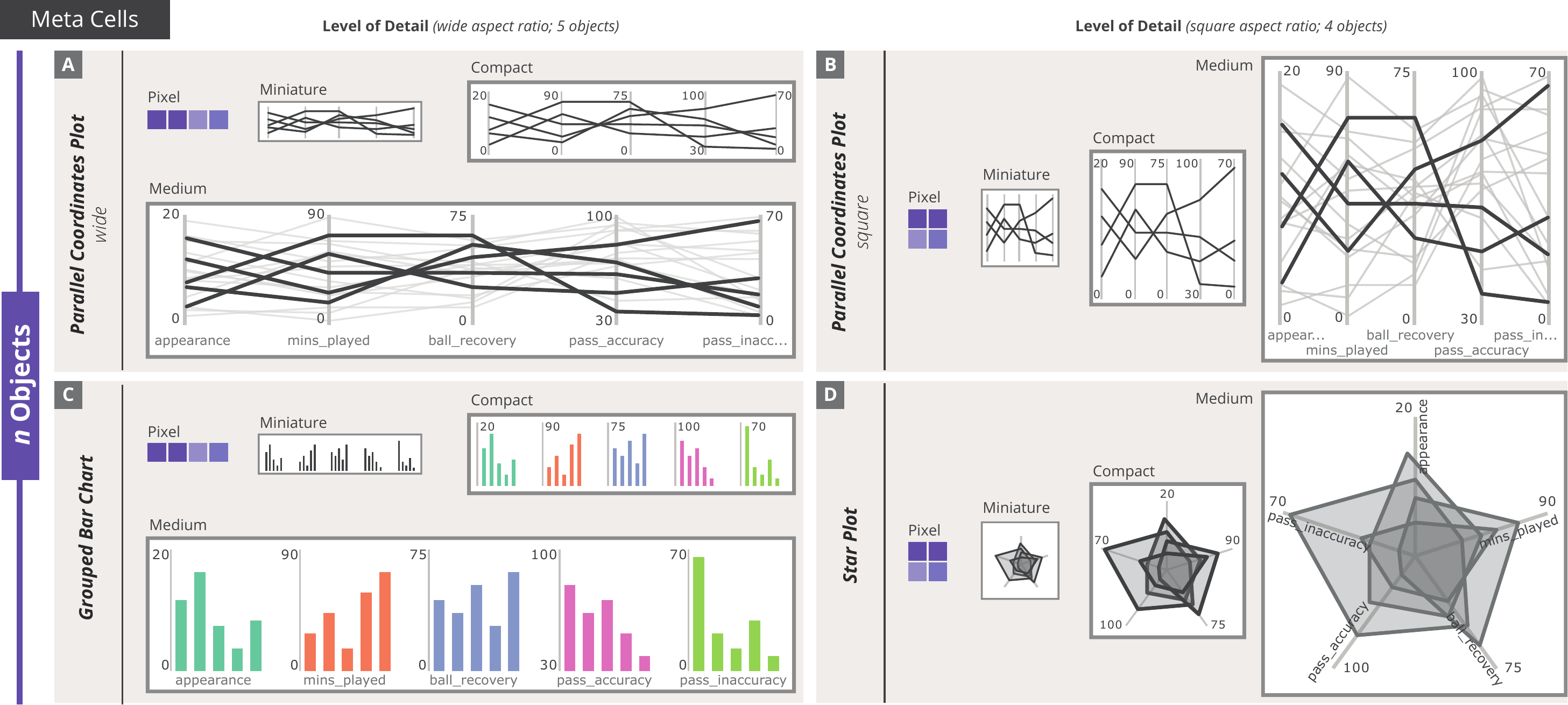}
    \vspace{-15pt}
	\caption{Visualizations in meta cells of size $i \times j$ must encode $n$ objects, either $n \leq i \cdot j$ edges or $n \leq i + j$ nodes. For example, when studying nodes, a $4 \times 1$ meta cell \protect\inlinefig{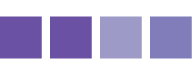} represents 4 column nodes plus one row node ($=$5 objects), while a $2 \times2 $ meta cell \protect\inlinefig{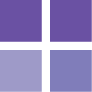} represents 2 column nodes plus 2 row nodes ($=$4 objects). In addition to different levels of detail, visualizations in meta cells have to take varying aspect ratios into account. \ccby}
	\vspace{-10pt}
	\label{fig:design_meta-cells}
\end{figure*}

This section proposes exemplary designs for multivariate visualizations in RMCs.
First, we focus on unit cells, for which the visualization has to encode either one or two objects primarily for \emph{object visibility}.
As suitable techniques, we consider bar charts and star plots for a single object as well as adaptations of them for representing and comparing two objects as illustrated in \autoref{fig:design_unit-cells}.
Second, we discuss visualization designs for meta cells, for which \emph{attribute visibility} is important.
Here, we consider parallel coordinates plots in addition to grouped bar charts, and star plots as indicated in \autoref{fig:design_meta-cells}.

\paragraph{Focusing on Details of a Single Object}

For a single object, the objective is to make its specific attribute values visible (R2). Bar charts are suitable for this purpose.
They already work well on the miniature size as bars are easy to distinguish and make good use of the available space (\autoref{fig:design_unit-cells}a).
At the compact size, it is possible to start showing labels (e.g., for the maximum), while at the medium size, all values and potentially the attributes can be labeled.

A downside of bar chart is that all attributes should be in the same or similar value range so that they can share the same axis.
Otherwise, certain attributes can be overemphasized if the same normalized axis is being used.
Alternatively, each bar can have its own axis, but these are difficult to incorporate on small sizes.
Another option is to configure the bars to not show absolute values but relative ones corresponding to the global min/max.
In both cases, however, interpreting the bars could be difficult as they would contradict typical conventions.

Another technique for representing a single object are star plots.
Similar to bar charts, star plots work well on miniature size thanks to their glyph-like appearance~\cite{Borgo2013GlyphbasedVisualization} (\autoref{fig:design_unit-cells}b).
The glyph-like character is particularly beneficial when multiple unit cells form a small-multiples arrangement.
Star plots make sense for three or more data attributes.
Each attribute has its own axis, avoiding issues with different attribute ranges.
However, the comparison of attribute values at the differently oriented axes can potentially be more demanding.
Value labels can be shown starting at the compact size, and axis labels make sense at the medium size.
Especially at smaller sizes, overlaps of labels with the plot lines and the axes are hard to avoid.

\paragraph{Comparing Two Objects in Detail}

In general, visual comparison of two objects can be supported by showing the two objects in parallel (superimposed or juxtaposed) or by computing and visualizing their difference directly~\cite{Gleicher2011VisualComparisonInformation}.
As before, bar charts and star plots can be used to show two objects at the same time.
Particularly useful for comparison are bar charts, where bars are grouped by attribute, and overlaid star plots~(\autoref{fig:design_unit-cells}c,d).
In both cases, the visual density is increased due to the additional graphical marks, which requires different responsive behavior.
For example, labels for the grouped bar charts become visible only at the compact size, as the miniature size already introduces the usage of different shades for the bars as a new detail.
For both grouped bar charts and overlaid star plots, it is not inherently clear which marks corresponds to which object (i.e., the node of the row or of the column).
This can be mitigated by establishing conventions.
For example, the bars corresponding to the row node can always be shown on the left, or its outlined polygon always be rendered on top.
Interactive coordinated highlighting further supports users in identifying data objects in RMCs~(see \autoref{sec:overview-to-detail-to-editing}).

In addition to showing two objects simultaneously, comparison tasks can also be supported by directly encoding the difference between the objects in a difference bar chart (\autoref{fig:design_unit-cells}e).
While this sacrifices the display of the actual attribute values, the comparison is simplified and the chart itself is cleaner with fewer marks being shown.
Thanks to the simpler design, difference bar charts work well in a small-multiples arrangement of unit cells.
The idea of encoding differences directly can also be expanded to star plots, in which case the polygonal shapes would encode the differences.

\paragraph{Inspecting Multiple Objects}

Meta cells provide a visual representation of a group of either nodes or edges.
In contrast to the designs discussed before, visualizations embedded into meta cells often divert from the typically square aspect ratio of their unit-cell counterparts.
In general, three aspect ratios of meta cells are relevant: a wide shape in horizontal orientation (landscape), a wide shape in vertical orientation (portrait), and an (almost) square shape.

Visualizations whose space demands grow mostly in only one direction work well with landscape and portrait, where different orientations can be supported by 90-degree rotation.
A prominent example are parallel coordinates plots (PCPs), which benefit from growing with the number of shown attributes or axes.
PCPs offer the necessary degree of flexibility to adapt to different aspect ratios as both the axes and the spacing in between are easy to adjust (\autoref{fig:design_meta-cells}a,b).
PCPs can support attribute visibility, which enables users to see how attribute values are distributed, whether attributes are correlated, or if there are any outliers.
At miniature size, no labels can be shown, while at compact size it gets possible to indicate minimum and maximum values per axis.
At medium size, axis labels can be displayed and the background can show the entire data set in a dimmed fashion to provide additional context.

Grouped bar charts and star plots also facilitate the inspection of multiple objects.
In a grouped bar chart, there are several adjunct groups of bars, one group for each attribute (\autoref{fig:design_meta-cells}c).
The grouping makes it possible to show individual axes per group at larger sizes.
Within a group, the number of bars corresponds to the number of objects.
Hence, the space demand for grouped bar charts primarily grows in only one direction as the number of attributes and objects increases.
This makes grouped bar charts suitable for landscape and portrait aspect ratios.
Star plots, on the other hand, become distorted for landscape and portrait.
They are better suited for square-shaped meta cells (\autoref{fig:design_meta-cells}d).
Analog to PCPs, star plots enable users to recognize attribute distributions and correlations as well as to compare specific objects.

As for any multivariate visualization, readability in meta cells decreases with a large number of marks due to over-plotting.
Yet, our focus+context RMCs are not meant to operate on larger data, but on subsets as defined by regions of interest.
Still, it is mandatory to support readability and object identification by means of interactive highlighting as described in \autoref{sec:overview-to-detail-to-editing}.

\subsubsection{Designs for Further Data Aspects}
\label{subsect:furtheraspects}

So far, we mainly illustrated RMCs for representing multivariate data aspects.
Yet, RMCs can also be employed to convey other data aspects, including structural, spatial, or temporal aspects of graphs.

While the adjacency part of the overview matrix already incorporates structural aspects, certain path-related analysis tasks are easier to carry out with node-link diagrams~\cite{Ghoniem2005ReadabilityGraphsUsing,  Okoe2019NodeLinkAdjacency}.
To combine the advantages of both, node-link diagrams can be embedded into meta cells.
They show the induced sub-graph corresponding to the set of nodes or the set of edges associated with the RoI.
The size and color of nodes as well as the stroke width of links can encode selected node and edge attributes.
Yet, this is mostly an option for larger meta cells.
Embedding a node-link diagram enables users to quickly check how certain patterns in the adjacency matrix look like in an arguably more intuitive representation.

Besides graph structure and multivariate attributes, a graph can have further facets, most prominently spatial and temporal dependencies~\cite{Hadlak2015SurveyMultifaceted}.
Provided that suitable visualizations for such additional facets exist, RMCs can generally be used to also embed them into the matrix.
For example, a meta cell could be extended to show a map underneath a node-link diagram and use a geographical layout rather than a force-directed layout.
Similarly, it would be possible to show nodes or edges along a time line. While these are first ideas for generalizing RMCs, concrete designs are left for future work.

\medskip

In summary, the RMC approach offers a great degree of flexibility in terms of \emph{what}, \emph{where}, and \emph{how} information is visualized. This naturally requires a high degree of interactivity as discussed in the next section.
\section{From Overview to Details to Editing}
\label{sec:overview-to-detail-to-editing}

\begin{figure*}[t]
    \centering
    \includegraphics[width=\textwidth]{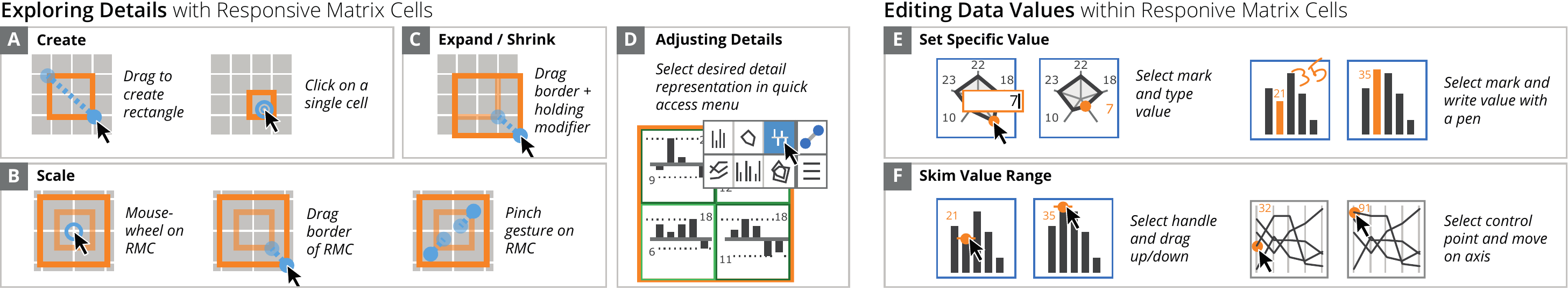}
    \vspace{-15pt}
    \caption{Interaction techniques for creating and configuring RMCs, and mechanisms for editing attribute values. \ccby}
    \vspace{-10pt}
    \label{fig:interactions}
\end{figure*}

To facilitate the dynamic use of RMCs as a data exploration and editing tool, a suitable interactive interface must be provided to the user.
In fact, our approach really lives from interaction.
Yet, the combination of focus+context and embedded visualizations makes the interface design a non-trivial endeavor.
On the one hand, interaction with the matrix must be possible on a global level (e.g., selecting attributes of interest).
On the other hand, users must be able to interact on a local level with the RMCs  (e.g., scaling RMCs) and the embedded visual representations (e.g., highlighting and editing data).
Careful design is necessary to obtain an easy-to-use and conflict-free interaction repertoire. 

The starting point for RMCs is that users spot something interesting in the overview matrix (R1).
Therefore, the analyst can initially configure the matrix on a global level by zooming and panning, selecting the attributes to be included in the similarity calculation, sorting rows and columns, and choosing appropriate color scales via a global menu.

Once the overview matrix has been set up so that interesting data features stand out, RMCs come into play to inspect and compare the surfaced features in detail (R2).
In the following, we discuss how analysts can create RMCs and configure the embedded visual representations.
Finally, we turn our attention to data editing by interactively manipulating graphical marks in the embedded visualizations (R3).

\subsection{Exploring Details with RMCs}

The primary steps for exploring details with RMCs are to create and configure RMCs in the first place, to adjust the embedded visual representations appropriately, and to link the gained insight across RMCs and the overview matrix.

\paragraph{Creating RMCs}

In order to create a new RMC, the analyst simply clicks and drags up a rectangular region of interest (RoI) covering the matrix cells to be studied in detail (\autoref{fig:interactions}a). 
A single-cell RMC is created with a single click.
As the user-specified RoIs are typically associated with some visual patterns being evident in the overview matrix (e.g., cluster of edges or group of very (dis)similar nodes), the creation process could be eased by offering automatic selection support that fits RMCs to such patterns~\cite{Yu16CAST}.

Upon creation, RMCs are initialized based on useful defaults.
Whether node or edge attributes will be shown (the what) depends on the triangular matrix part where the RoI is created.
By default, meta cells will be created (the where). 
To generate a small-multiples arrangement of unit cells, a modifier key (e.g., shift) can be held while selecting the RoI. 
For the embedded visualization (the how), we consider bar charts as a suitable default.
All these default settings can be subject to interactive adjustment via a local menu as explained later. 

\paragraph{Scaling RMCs}

A major advantage of RMCs is their flexible level of detail (LoD), which is coupled to their scaling level.
On creation, RMCs are automatically scaled up from the pixel to the miniature level revealing initial details in the embedded visualization.
The analyst can increase the LoD further by local zooming, for example, using the mouse wheel, dragging the RMC borders, or performing a pinch gesture (\autoref{fig:interactions}b).
The additional space required for enlarging RMCs is obtained by shrinking rows and columns outside of RMCs uniformly like in bifocal views~\cite{Apperley1982BifocalDisplayTechnique}.
To deal with the issue of varying aspect ratios, the zooming can happen either uniformly in $x$ and $y$ directions or be restricted to only $x$ or $y$ direction.
Upon zooming, responsiveness sets in and RMCs are automatically enhanced with additional information and richer visual encodings.
These make it easier for the analysts to read the visual representation and understand details better.

% As selecting and scaling are often used in combination, it is helpful to support both within a single continuous interaction. Specifically, this can be supported through a ``create--dwell--scale'' scheme: when drawing a rectangular and holding for a short amount of time, the selection is automatically confirmed and the mode switched to scaling by dragging its border.
% This avoids the need to target and click the border again, before the desired scaling can be applied.
% Such a combination of interactions is also known as phrasing.

\paragraph{Expanding, Shrinking, and Dismissing RMCs}

After first conclusions have been drawn from an RMC, the analyst's interest might change.
This can result in the need to adapt the region covered by RMCs, that is, to expand or reduce it by adding or removing cells.
This can be supported by dragging borders similar to scaling up RMCs, but while activating another modifier (\autoref{fig:interactions}c).

Once an RMC's details have been studied conclusively, the RMC can be dismissed.
This is as easy as triggering a shortcut key (e.g., delete) or a designated mouse button. 
A global reset function can be used to dismiss all RMCs altogether and reset the overview matrix.

\paragraph{Adjusting the Display of Details}

To facilitate the in-depth exploration of details, RMCs provide an in-place menu interface (\autoref{fig:interactions}d) for adjusting what (nodes or edges), where (unit or meta cells), and how (embedded visualization) details are made visible.
When switching nodes and edges, RMCs are automatically transitioned from one half of the matrix to their corresponding position in the other half.
Switching the cell type results in either merging a set of unit cells into a meta cell or splitting up a meta cell into several unit cells.
Switching the embedded visualization simply replaces the visual representation in an RMC.
To allow analysts to quickly switch back and forth between the different options, the menu stays open until a suitable configuration has been found and the menu is closed explicitly.

As we consider altering the visualization (the how) to be a frequent operation during the data exploration, additional shortcuts are provided.
The arrow keys can be used to select different visualizations and layout variants, the space bar toggles between unit and meta cells, and the tab key switches between node and edge attributes.

Exploring details typically involves further adjustments of visual representations, for example, reordering axes, selecting attributes, changing scales, and so on.
While it is standard to carry out such interactions directly within the visualization, this is impractical for our space-constrained RMCs.
Instead, it makes sense to offload further adjustments to external controls or the menu.

\paragraph{Linking Details and Overview}

A coordinated highlighting is indispensable to support analysts in linking the details provided in one RMC to the overview matrix and the details in other RMCs.
In general, hovering graphical marks in RMCs results in highlighting all other marks being associated with the same node or edge.
For example, hovering a node in an embedded node-link diagram results in highlighting all corresponding marks in all other RMCs and in emphasizing the corresponding row and column labels in the overview matrix (and vice versa).

\begin{figure*}[t]
    \centering
    \includegraphics[width=\textwidth]{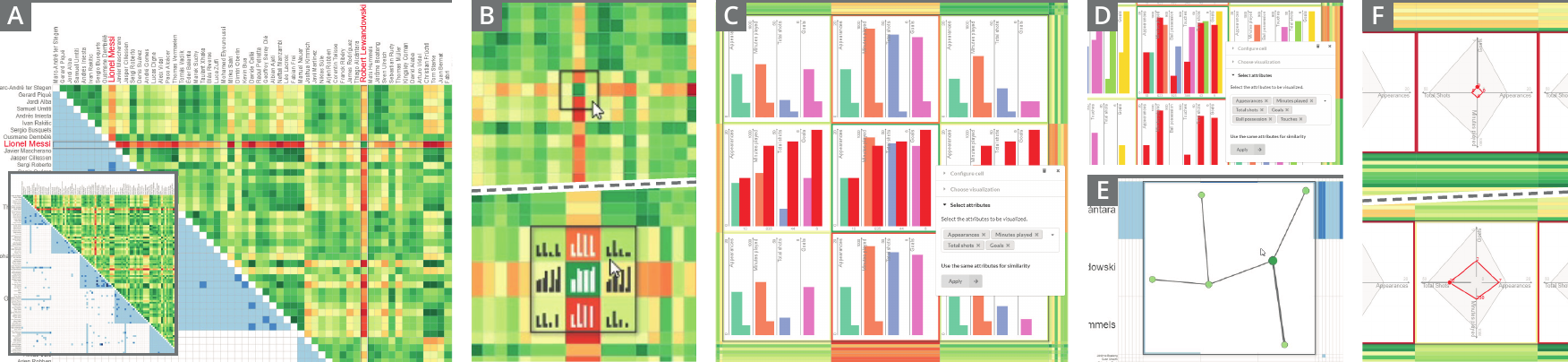}
    \vspace{-15pt}
    \caption{From overview (A), to exploring multivariate (B--D) and structural (E) details, to editing data values (F) of soccer player data. \ccby}
    \label{fig:walkthrough}
    \vspace{-15pt}
\end{figure*}

\subsection{Editing Data Values within RMCs}

During an in-depth analysis of a multivariate graph, it can be desirable or even necessary to shift from data exploration to data editing.
This shift can be motivated by the need of either correcting erroneous data or observing the influence of an attribute on the overall graph.
The first case corresponds to Baudel's direct manipulation principle, where data values are edited directly within the visualization~\cite{Baudel2006InformationVisualizationDirect}.
The second case addresses what Spence coined \emph{what-if} analyses, which can help users understand the interplay of different values% in the data
~\cite{Spence2001InformationVisualization}.
In both situations, the edits are supposed to be immediately visible in the visualization.

In general, an edit operation can target the graph structure (add/remove nodes/edges) or the associated attribute values (update)~\cite{Gladisch2015MappingTasksInteractions}.
The literature already offers several strategies for editing structural aspects using matrix representations~\cite{Gladisch2014SemiAutomaticEditing,Gladisch2015usingMatrixVisualizations, Kister2017GraSpCombiningSpatially}.
Therefore, our interest primarily regards the editing of attribute values.

Depending on the user's goal (correcting error or what-if analyses), editing can mean plainly setting a specific value or involve skimming a range of potential values before one value is eventually set.
A specific value is easy to set by entering it via keyboard or, where pen input is available, via handwriting.
Both ways are illustrated in \autoref{fig:interactions}e.

For what-if analyses, entering many values in such a discrete fashion is impractical.
Instead, it must be possible to quickly check a range of values while observing the resulting changes in the visualization.
This is facilitated by continuous drag gestures where users move the data-encoding marks directly within RMCs.
To this end, interaction handles become available as soon as RMCs are sufficiently large to allow for a reasonable range of movement so that edits can be performed with an acceptable precision.
For most of the previously discussed visualizations, this starts to be doable at the compact size.
To support what-if questions like \emph{what if the attribute value would be similar to another object}, it can be beneficial to 'snap' graphical marks being dragged to other data objects.

\autoref{fig:interactions}f illustrates how editing works in bar charts and parallel coordinates.
In a bar chart, the upper end of a bar can be dragged up and down in order to update the underlying attribute value.
In parallel coordinates (and also in star plots), the control points of the polylines can be dragged for editing.

\medskip

Taken together, RMCs offer a versatile interaction repertoire supporting users in conducting comprehensive analyses of multivariate graphs without switching to external tools.
How RMCs and the offered interactions can be applied to a specific data set will be shown next.
\section{Applying Responsive Matrix Cells}

We implemented the RMC approach in a web-based prototype using the native canvas API for rendering, the D3 library for computing force-directed layouts, and the chroma.js library for color coding. The GUI consists of a mix of SemanticUI and custom controls.
The prototype supports all key concepts via mouse and keyboard, including creating unit and meta cells, scaling them up, changing the visualizations, and editing attribute values.
Except for difference bar charts and spatio-temporal visualizations, all visualizations discussed in \autoref{subsect:rmc-visualizations} are implemented.
The prototype is publicly available~\cite{Tominski20RMCProto}.

Here, we demonstrate the feasibility of our approach by means of a walk-through for the use case of exploring and editing real-world soccer data as well as by reporting on early user feedback. 
For future work, the walk-through could also be used as a basis for defining tasks of a more elaborate user study.

\subsection{Data \& Task}
\label{subsect:prototype}

As an example data set, we used a graph of soccer players from the 2017/18 Champions League season.
The graph consists of 95 players, the nodes of the graph.
The players are characterized by up to 39 quantitative data attributes, including general stats (e.g., minutes played), defensive figures (e.g., balls recovered, interceptions), and offensive qualities (e.g., shots on goal, goals scored).
Not all players have values for all attributes.
While this is partly due to different player types (e.g., goal keeper can have special attributes), some players are actually lacking correct attribute values.

There are 1046 edges in the graph.
An edge represents co-occurrences of two players, that is, if and how often the two players have played for the same club during their career.
The edge weight corresponds to the number of shared clubs. There are no further edge attributes present in the data set.

For our walk-through, we assume the goal of an analyst is to identify match-deciding players and compare them with each other.
In particular, this includes (i) the exploration of details of an a priori unknown sub-graph and (ii) the correction of found errors within this sub-graph.

\subsection{Walk-through: Exploring \& Editing Soccer Players}

Initially, the overview matrix shows the 95 players from 5 different clubs (\autoref{fig:walkthrough}a).
Consequently, the matrix comprises 9,025 cells in total, of which 4,465 are available for showing the edges, 4,465 for showing the node similarities, and 95 cells in the diagonal.
The nodes (rows and columns) are initially ordered by club, with other orders being available.
The similarity part is calculated on the fly based on a user-specified set of attributes.
In order to focus on match-deciding players, the analyst chooses the minutes played, appearances, total number of shots, and scored goals as the attributes for the similarity calculation.
Then, the similarity part of the matrix clearly shows two players standing out as lines with many reddish cells, indicating dissimilarity to all other players in the data set (\autoref{fig:walkthrough}a).
The players standing out are Lionel Messi (LM) and Robert Lewandowski (RL).
While being dissimilar to other players, they are quite similar to each other as indicated by a green cell exactly where the two reddish lines cross.

In order to study and compare both players in detail, the analyst selects a sub-graph around the green cell representing the high similarity of LM and RL (\autoref{fig:walkthrough}b).
Holding down the shift key while making the selection results in a $3 \times 3$ small-multiple arrangement of unit cells being created.
The unit cells are automatically enlarged and populated with bar charts.
The analyst zooms in and focuses on the cell representing LM and RL, which can still be recognized easily by the maintained green background (\autoref{fig:walkthrough}b).
Already at this small size, it can be seen that LM and RL have indeed similar attribute values as the bars in each of the four bar groups have similar lengths.

Scaling up further will add axes labels and colors for better distinguishing the four bar groups (\autoref{fig:walkthrough}c).
Moreover, red bars highlight the data associated with LM in all unit cells.
In the central unit cell, the analyst can see that LM and RL have the exact same number of appearances, and the number of minutes played and the number of total shots are quite similar.
The detailed representation also enables the analyst to infer that the shot-to-goal ratio is roughly the same for RL and LM (43:5 vs. 44:6), although RL has played more minutes.

To further investigate the two players, the analyst opens the quick access menu (\autoref{fig:walkthrough}c) and fetches the attributes ball possession and number of touches as additional details to the chart.
Now, significant differences become visible (\autoref{fig:walkthrough}d): LM has twice as much ball possession and twice as many touches as RL.
The discovered similarities and differences suggest two different playing styles.
LM could have a greater impact on the game, as he not only creates many decisive game situations, but also completes them successfully.
In comparison, RL seems to be less active in creating situations with the ball, but can decide games with good positional play and ultimately scoring.

As these two styles complement each other well, it would be interesting to know if the two players have ever played for the same club.
To answer this question, the analyst switches the RMC to the adjacency part of the matrix, where the cells encode the number of clubs shared.
The overview matrix suggests that a few players have played together in more than one club.
RL and LM have never played together, as indicated by an empty white cell in the adjacency part.
To get a more intuitive representation, the analyst switches to a meta cell with an embedded node-link diagram (\autoref{fig:walkthrough}e).
Through interactive highlighting, the analyst can find out that one path between RL and LM exists in the sub-graph via Thiago and Demb\'el\'e.

Besides decisive players that are almost always on the field, there are also key substitute players who enter a match in critical situations.
The analyst remembers three such potential substitutes who appeared as particularly dissimilar to LM earlier during the exploration.
To confirm that these players have high offensive skills, had a couple of appearances, but played only a few minutes, the analyst goes back to the similarity part of the matrix and creates unit cells with embedded star plots (\autoref{fig:walkthrough}f).
The star plots reveal that several attribute values are missing, and hence, the players are not necessarily dissimilar to LM.

The analyst decides to fix these missing values.
After zooming to the medium size, handles appear at those marks in the visualization where edits are possible.
By simply dragging a handle, the analyst can change a data value and update it to a correct one (\autoref{fig:walkthrough}f).
During the editing, the star plot is updated on the fly, as are all other visualizations depending on the edited value, including the similarity part of the matrix.
After a correct value has been set, the value is committed to the data and the analysis session can be continued seamlessly by discarding the existing RMC and defining new ones in other parts of the matrix.

\subsection{Preliminary User Feedback}

In order to receive early user feedback on our approach, we invited 4 researchers (2 interaction experts, 2 visualization experts, all PhD-level) for guided hands-on sessions.
After an introduction and short demonstration, participants (P1-4) were asked to interact with the prototype and test the usability of its different functionalities.
Sessions were conducted remotely by two investigators via video chat with screen-sharing and lasted ca.\ 1 hour.
Overall, participants were very positive and attested the implementation a high quality.
While all agreed that initial training is required to understand both data set and visualization approach, participants adapted to the interface quickly and used all techniques without larger issues.
Interestingly, while we did not instruct for, all participants started to reason about the data, but with different approaches and focus.
For example, P3 started exploring possible matrix sorting, P1 looked into node similarities, and P4 focused on the relations between similarity and adjacency.

Both unit cells and meta cells were considered helpful to understand why nodes are (dis)similar, but we could observe that unit cells required more time to be properly read---likely since attribute labels are only shown on higher zoom levels.
As P1 and P4 used global zoom more intensively, they noted that the node labels were quickly becoming invisible, as they are only placed outside of the matrix.
Showing the labels additionally around an RMC could avoid this.
The highlighting mechanisms were considered useful with few suggestions for improvements, e.g., permanent highlights for one or more nodes (P2, P4), or highlights of attribute axes when hovering the labels in the sidebar or context menu (P2). 
All participants found the editing very useful, particularly for understanding the influence of certain attributes on the similarity measure.
However, while working with larger RMCs at a high LoD, P2 and P4 noted that due to the stronger distortion the edit effects are getting harder to observe in the overall matrix.
Simplifying editing on lower LoDs could mitigate this issue.
With most mechanisms working smoothly, ideas for further functionalities were also proposed, e.g., to allow filtering of nodes within the RMCs (P3).

\medskip

The above walk-through as well as the user feedback provide a first indication of the utility of RMCs.
As we cannot show further details here, we refer to the supplementary video, which much better illustrates the dynamic nature of the RMC approach.
Walk-through, user feedback, and supplementary video suggest that RMCs accomplish what they set out to achieve: They support a seamless analysis workflow from an overview to details to editing without resorting to external tools.

\section{Discussion}

Our approach is based on a non-trivial interplay of several visual (e.g., matrix, charts), interactive (e.g., focus+context, highlighting), and automatic (e.g., similarity computation, responsiveness) components.
Next, we further discuss limitations and possible extensions of RMCs.

\paragraph{Facets of Responsiveness}

In the visualization domain, the notion of responsiveness is typically focused on adapting a visualization based on space constraints, e.g., adapting it to the small screen of a mobile device~\cite{Andrews2017ResponsiveDataVisualisation, Hoffswell20Responsive}.
In contrast, the responsiveness of our RMCs is not limited to size-based adaptations, but also includes the consideration of the underlying data (similar to semantic zooming), and the analyst's tasks.
In the future, this should be complemented by the used input modality (e.g., mouse, touch, pen, natural language).
At the same time, the question remains how visualizations should respond to these different facets, i.e., how they can be adapted in a useful way or when the representation should be set to a different visualization type.
For future work, one goal would be to investigate what ``useful'' adaptations are and what users would expect as responsive behavior.

\paragraph{Automation Desiderata}

Another goal could be to automate certain aspects of RMCs.
To reduce the overall interaction costs~\cite{Lam08InteractionCosts}, the automation of recurring adaptation patterns can be considered.
Such patterns can be sequences of changes (first star plots, then bar charts, then node-link) or RoI-specific changes (unit cells for smaller RoIs, meta cells for larger ones).
However, interaction patterns like these are typically highly application- and user-dependent, which prohibits hard-wiring automated adjustments into RMCs.
Instead, machine learning methods could infer automatable adjustments from previous user interactions~\cite{Brown14LearningInteractions, Endert17MachineLearning, Ottley19FollowClicks}.
Where this is not possible (e.g., due to a too small user base), templates could define potentially helpful rules for automatic adjustments~\cite{Tominski11EventBased}.

\paragraph{Embedded Visualizations}

An important part of RMCs is to have a suitable and diverse set of visualizations.
While we already provide variations of bar charts, star plots, and PCPs, this collection can be extended in the future.
Particularly interesting are tailored visualizations that work well for specific constellations.
For example, further glyph-like visualizations can be effective for small unit cells, scatter plots could show correlations between two attributes in meta cells, miniature maps would be helpful for geo-spatial networks, and horizon graphs could be applied to temporal data attributes.
Besides considering additional visualizations, it could also make sense to think about combining or overlaying them within RMCs.

\paragraph{Extended Editing Facilities}

Our approach eases the hurdles of data editing by enabling it directly within the RMCs.
This means that exploration and editing can take place in the very same context.
Yet, the editing could be further improved.
For example, any editing is ideally supported through a history mechanism for undoing and redoing edits and capturing insights~\cite{Kreuseler04History, Nancel14History, Mathisen2019InsideInsightsIntegratingData}.
Such provenance features are particularly useful for extensive analysis sessions.
For the interaction itself, interactive surfaces such as tablets or digital whiteboards are interesting environments for editing.
Specifically, the additional input modalities of these environments, e.g., touch~\cite{Horak2018WhenDavidMeets, Sadana2014DesigningImplementingInteractive}, pen~\cite{Frisch2009Investigatingmultitouch, Romat2019ActiveInkThInkingData}, or speech~\cite{Srinivasan18Orko}, can potentially simplify edit operations and improve precision at lower LoDs.
For example, in order to update an attribute value, the new value could simply be spoken, written with a pen, or indicated by ''slicing`` a bar at a certain height via touch.

\paragraph{Matrix Scalability}

A general challenge for multivariate graphs is scalability~\cite{Jankun-Kelly2014ScalabilityConsiderationsMultivariate}.
When using a matrix visualization, larger graphs with more than a couple of 100 nodes are getting difficult to handle, both conceptually and implementation-wise.
The first aspect can potentially be tackled with the help of existing approaches such as Graph Sketches~\cite{Abello02GigaGraphs} or NodeTrix~\cite{Henry2007NodeTrixHybridVisualization}.
Implementation-wise, due to the quadratic complexity of a matrix and the additional elements added with the RMCs, our prototype is currently limited to rendering roughly 150 nodes.
With more sophisticated GPU-based rendering, this limit could be expanded, but only by a constant factor.

% \paragraph{Provenance}
% Similar to other visualization solutions, an analyst is focusing on many different details and aspects over the course of an analysis session.
% Within our RMC approach, this means that many open-scale-close cycles of ROIs and RMCs are performed.
% In situations where an analyst wants to circle back to a previous investigated aspect, it can get difficult to remember the exact configuration of the used RMC.
% Here, an enhanced support of provenance (e.g., history of steps, reopening of previous ROIs, undo of configurations) can be highly beneficial.
% This can be an important addition when developing the here presented concepts further into a complete visualization system.
% % Notably, our general workflow approach of going from overview to detail is only one possibility; 
% % different approach than from overview to detail possible: cf.\ Elzen et al.~\cite{Elzen2014MultivariateNetworkExploration} (from detail to overview via selections and aggregations)

\paragraph{Formal User Studies}

For developing our approach, we followed an iterative design process with input from both visualization and HCI experts.
Such expert input has proven to be very valuable in similar contexts~\cite{Nobre2020EvaluatingMultivariateNetwork}.
Here, we demonstrated by means of a walk-through how our concepts come into play during an analysis session and provided insights from a preliminary user feedback session.
A supplementary video is available to illustrate the dynamic nature of RMCs.
In the future, it would be interesting to conduct formal user studies to investigate the RMCs in more depth.
However, studies in the context of integrating various visualization views are non-trivial~\cite{Plumlee06Overview, Nekrasovski06EvalRubbersheet}.
In our case, the provided editing mechanisms would even add to the study complexity.
In addition to usability studies, comparative studies could contrast our integrated approach to an alternative with multiple juxtaposed views.
In order to foster further investigations into our RMC approach of any kind, we made our prototype freely available.

\paragraph{RMCs beyond Multivariate Graphs}

Finally, it can be promising to investigate the potential of RMCs in other contexts.
RMCs can naturally be applied to any tabular arrangement of cells where cells contain information worth being elaborated with additional details.
A classic example is the TableLens~\cite{Rao1994TableLensMerging} and variants of it~\cite{McLachlan08LiveRAC, John2008VisualAnalyticalExtensions}.
In the future, it would also be interesting to explore the applicability of RMCs in the context of irregular arrangements, for example, if and how RMCs can be applied to maps or map-like visualizations~\cite{Hograefer20MapLike}.

\paragraph{Limitations}

Our RMC approach shares conceptual limitations of the base concepts we combine.
In general, table and matrix representations depend very much on an appropriate ordering of rows and columns~\cite{Behrisch16MatrixOrdering}.
In our context, this is particularly important because only adjacent cells can be turned into RMCs to reveal details.
Addressing this issue, we integrate options to sort the matrix with respect to graph characteristics (e.g., degree, pre-defined clusters), data attributes, or node similarity.
Like many focus+context techniques, our RMC approach distorts the basic visualization layout, which can make it more difficult to follow rows and columns in the matrix.
We addressed this by providing coordinated highlighting and a cross-hair cursor that spans the entire matrix.
Moreover, thanks to the scalable display of information in RMCs, they can also be applied at miniature or compact sizes, which require only minimal distortion.
Finally, it should be acknowledged that our approach currently is tailored to un-directed multivariate graphs.
For directed graphs, more complex encodings indicating the direction in the adjacency part must be considered, or the half of the matrix encoding attribute-related aspects must be omitted.
For the latter case, one could place a data table directly next to the matrix~\cite{Berger2019VisuallyExploringRelations}, where the matrix shows the graph structure and the table visualizes node attributes.
Both matrix and table could then offer the option to embed RMCs for detailed exploration and editing.

\section{Conclusion}
Exploring and editing multivariate graphs is a complex task that requires considering the structure and the multivariate aspects of nodes and edges on both an overview and detail level. 
With our responsive matrix cells approach, we enable such analyses by providing embedded visualizations in user-defined focus regions within a matrix visualization.
The RMCs allow users to flexibly define what, where, and how detail information is to be represented.
Consequently, our approach facilitates both exploring and editing those details, which are otherwise obscured in the complexity of a multivariate graph.
RMCs enable workflows following and extending Shneiderman's information seeking mantra~\cite{Shneiderman1996}: Start with an overview, then zoom into details of interest, explore these details from different perspectives, and edit them if required.
The versatility of RMCs is formed by the well-known concepts of focus+context, semantic zooming, and direct manipulation, combined with the emerging ideas of responsive visualizations.
As illustrated by our walk-through and the user feedback, we believe that this combination is a promising solution for analyzing complex multivariate graphs in general and is worth to be pursued further, not only in an integrated fashion in matrix visualizations but also beyond.
\acknowledgments{
We thank Tatiana von Landesberger for her valuable feedback and Eva Goebel for assisting us during the early drafts.
The soccer data set was prepared in collaboration with Mohammad Chegini.
This work was supported by the DFG grant 214484876 (\emph{GEMS 2.0}) and grant 389792660 as part of TRR~248 (see \href{https://perspicuous-computing.science}{perspicuous-computing.science}).
}

% \input{content/notes.tex}
% \input{content/story.tex}
% \newpage

%\bibliographystyle{abbrv}
% \bibliographystyle{abbrv-doi}
%\bibliographystyle{abbrv-doi-narrow}
\bibliographystyle{abbrv-doi-hyperref}

\bibliography{refs}
\end{document}